\documentclass[letter]{aa} 

\usepackage{graphicx}
\usepackage{txfonts}
\newcommand{\doce}{\mbox{$^{12}$CO}}

\newcommand{\trece}{\mbox{$^{13}$CO}}

\newcommand{\jsc}{\mbox{$J$=6$-$5}}

\newcommand{\jtd}{\mbox{$J$=3$-$2}}
\newcommand{\jdu}{\mbox{$J$=2$-$1}}
\newcommand{\juc}{\mbox{$J$=1$-$0}}

\newcommand{\kms}{\mbox{km\,s$^{-1}$}}

\newcommand{\ms}{\mbox{$M_{\mbox{\sun}}$}}

\newcommand{\lsim}{\raisebox{-.4ex}{$\stackrel{\sf <}{\scriptstyle\sf \sim}$}}
\newcommand{\gsim}{\raisebox{-.4ex}{$\stackrel{\sf >}{\scriptstyle\sf \sim}$}}

\newcommand{\farcss}{\mbox{\rlap{.}$''$}}
\begin{document}

   \title{ALMA observations of the Red Rectangle, a preliminary analysis}

   \author{V. Bujarrabal
          \inst{1}
          \and
          A. Castro-Carrizo\inst{2} 
          \and J. Alcolea\inst{3} \and H. Van
          Winckel\inst{4} \and C. S\'anchez Contreras\inst{5} \and
M.\ Santander-Garc\'{i}a\inst{3,6} \and R. Neri\inst{2} 
\and R. Lucas\inst{7}
          }

   \institute{             Observatorio Astron\'omico Nacional (OAN-IGN),
              Apartado 112, E-28803 Alcal\'a de Henares, Spain\\
              \email{v.bujarrabal@oan.es}
\and 
 Institut de Radioastronomie Millim\'etrique, 300 rue de la Piscine,
 38406, Saint Martin d'H\`eres, France  
\and
             Observatorio Astron\'omico Nacional (OAN-IGN),
             C/ Alfonso XII, 3, E-28014 Madrid, Spain
\and
Instituut voor Sterrenkunde, K.U.Leuven, Celestijnenlaan 200B, 3001
Leuven, Belgium
\and 
Centro de Astrobiolog\'{\i}a (CSIC-INTA), ESAC Campus, E-28691
Villanueva de la Ca\~nada, Madrid, Spain 
\and
Centro de Astrobiolog\'{\i}a (CSIC-INTA), Ctra. M-108, km. 4,
E-28850 Torrej\'on de Ardoz, Madrid, Spain 
\and 
UJF-Grenoble1/CNRS-INSU, Inst.\ de Plan\'etologie et d'Astrophysique 
de Grenoble (IPAG) UMR 5274, Grenoble, F-38041, France
           }

   \date{July 2013, accepted}

  \abstract
   {}
{We aim to study equatorial disks in
   rotation and axial outflows in post-AGB objects, as to disclose the
   formation and shaping mechanisms in planetary nebulae. So far, both disks
   and outflows had not been observed simultaneously.}
{We have obtained high-quality ALMA observations of \doce\ and
  \trece\ \jtd\ and \doce\ \jsc\ line emission in the Red Rectangle,
  the only post-AGB/protoplanetary object in which a disk in rotation
  has been mapped up to date. }
{These observations provide an unprecedented description of the
  complex structure of this source. Together with an equatorial disk in
  rotation, we find a low-velocity outflow that occupies more or less
  the region placed between the disk and the optical X-shaped nebula. 
From our observations 
  and preliminary modeling of the data, we confirm the previously known
  properties of the disk and obtain a first description of the
  structure, dynamics, and physical conditions of the outflow. 
}
   {}


   \maketitle
%

\section{Introduction}

Many protoplanetary nebulae (PPNe) show very massive ($\sim$ 0.1 \ms)
and fast (30--200 \kms) bipolar outflows, which are thought to be
crucial in the formation of planetary nebulae; see e.g.\ Bujarrabal et
al.\ (2001), Balick \& Frank (2004). These outflows carry too much
linear momentum to be powered by momentum transfer from stellar
photons, and the presence of disks rotating around the central stars
and reaccretion from them are often postulated to explain the nebular
dynamics (e.g.\ Soker 2001, Frank \& Blackman 2004). However, only one
of these putative disks has been well identified up to date, by means
of interferometric mm-wave maps of \doce\ \juc\ and \jdu\ emission in
the Red Rectangle (Bujarrabal et al.\ 2005). The Red Rectangle is a
well studied PPN and, curiously, one of the post-AGB objects that do
not show massive and very fast outflows.
It consists of the equatorial rotating disk plus a spectacular
axisymmetric nebula seen in the visible, surrounding a double stellar
system (Men'shchikov et al.\ 2002, Cohen et al.\ 2004).  It also shows
some properties, as a NIR excess indicative of hot dust kept close to
the stellar system, that suggest the presence of a very compact disk
(e.g.\ Van Winckel 2003).

Recent single-dish observations of \doce\ and \trece\ mm-wave emission
in a sample of similar post-AGB stars (close binary stars with low-mass
nebulae and indications of compact disks) systematically yielded
characteristic line profiles, with a prominent single or double peak
and moderate-velocity wings, which are strikingly similar to those of
the Red Rectangle (Bujarrabal et al.\ 2013, Paper II).
Profiles of this kind are also found in disks around young stars
(notably T Tauri variables), and have been proven, both from theoretical
and observational grounds, to be very reliable indicators of rotating
disks. 

Remarkably, gas in slow expansion was also detected in some of these
post-AGB objects. CO maps of 89 Her show an extended component
expanding at about 5 \kms, whose emission dominates the line wings
(Bujarrabal et al.\ 2007); this source also shows a barely-resolved
central condensation that is probably in rotation. From careful
modeling of Herschel observations of high-$J$ CO lines in the Red
Rectangle, Bujarrabal \& Alcolea (2013, Paper I) deduced that a
(probably bipolar) component in expansion should significantly
contribute to the high-$J$ line wings in this source.  But the spatial
resolution of those observations was poor and the existence of such a
component remained very uncertain.  Indications of molecule-rich
outflows were also found for other similar sources, in which the
disk-like CO profiles show relatively strong line wings (Paper II).

All these nebulae, including the Red Rectangle, show low values of the
total mass (\lsim 10$^{-2}$ \ms) and expansion velocity (5--10
\kms), contrary to the case of the massive PPNe mentioned before, and
may represent a different post-AGB evolutionary regime.  The Red
Rectangle is the prototype of this important class of objects and the
best target to study the simultaneous appearance of rotating disks and
bipolar outflows, a basic input for understanding the formation and
shaping of planetary nebulae.

   \begin{figure*}
   \centering
   \includegraphics[width=17.3cm]{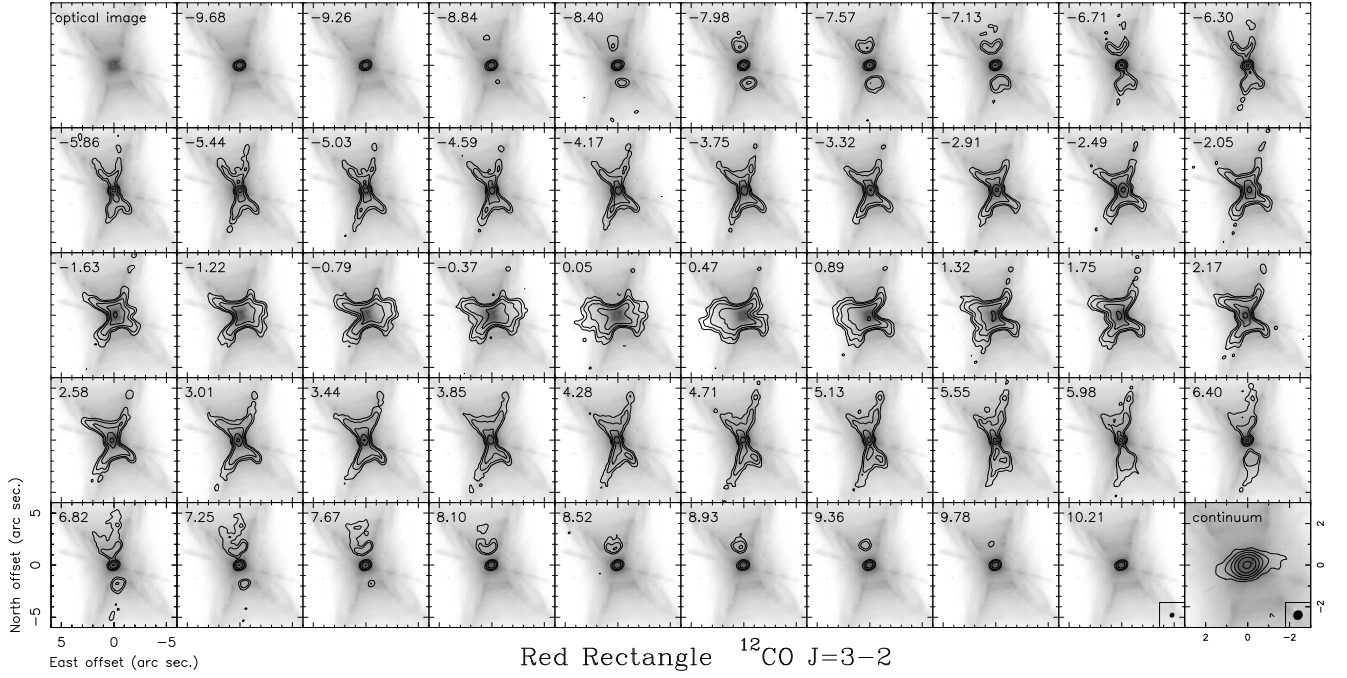}
      \caption{ALMA \doce\ \jtd\ observations of the Red Rectangle. The
        maps are centered at J2000 coordinates 06:19:58.219,
        -10:38:14.71. The LSR velocity is indicated in the upper-left
        corners. The contour spacing is logarithmic: -0.03, 0.02, 0.06,
        0.18, 0.54, and 1.62 Jy/beam (equivalent to 0.67, 2.0, 6.1,
        18.2 and 54.5 K). The HST optical image is also displayed and
        the last panel shows the 0.85mm continuum image zoomed by a
        factor 2 and with contours $\pm$0.0015, 0.0045, 0.0135, 0.0405,
        0.1215, and 0.3645 Jy/beam. The inserts in two last panels show
        the beam width. }
         \label{}
   \end{figure*}

   \begin{figure*}
   \centering
   \includegraphics[width=17.3cm]{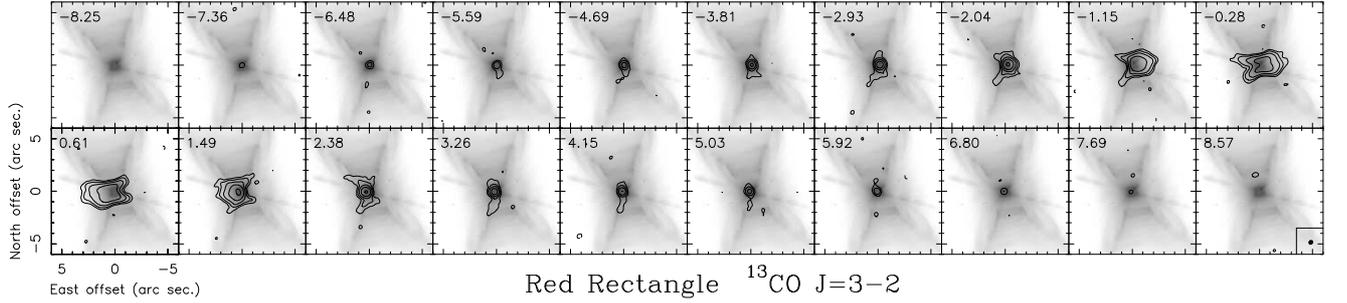}
      \caption{Same as Fig.\ 1 for \trece\ \jtd.  For this
        weak line, the continuum has been subtracted and the
        spectral resolution has been degraded to 0.84
        \kms. The contour spacing is logarithmic:  
$\pm$0.015, 0.045,
        0.135, 0.405, and 1.215 Jy/beam ($\pm$0.51, 1.53, 4.59, 13.8, and
        41.3\,K).  }
         \label{}
   \end{figure*}

   \begin{figure*}
   \centering
   \includegraphics[width=17.3cm]{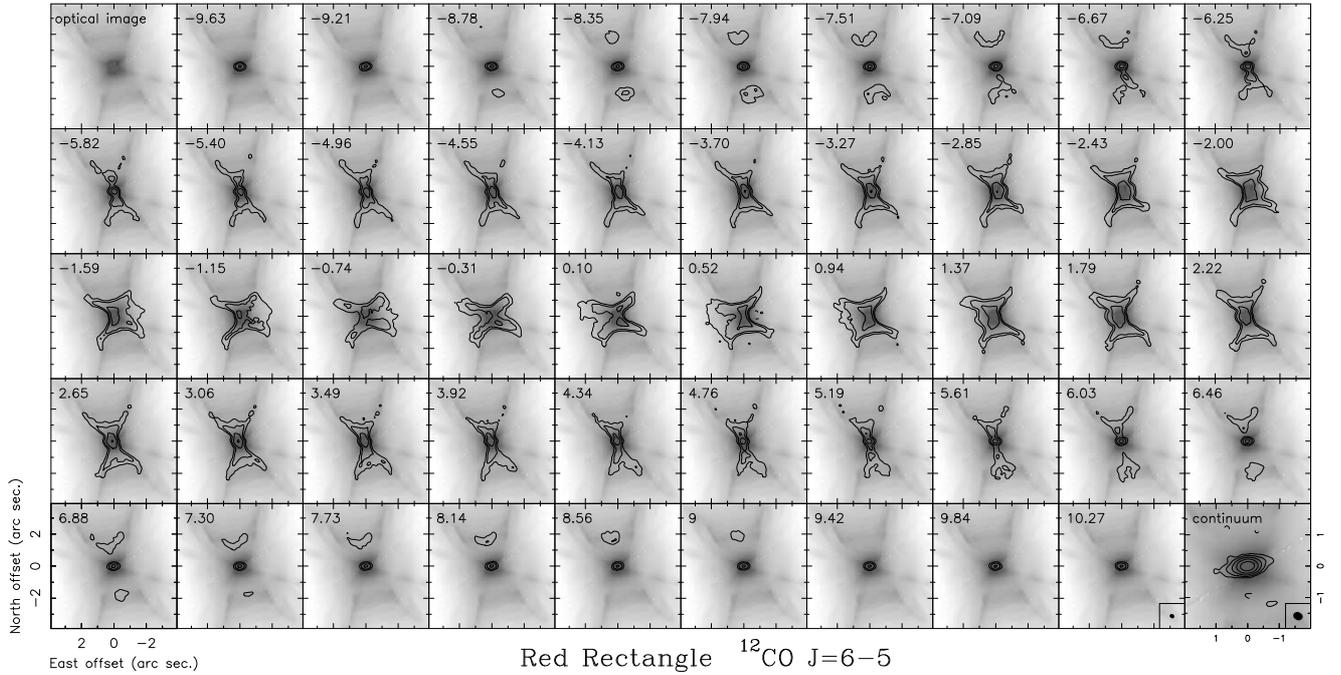}
      \caption{Same as Fig.\ 1 but for \doce\ \jsc. The contour spacing
        is logarithmic: 
        -0.3, 0.2, 0.6, 1.8, and 5.4 Jy/beam (6.7, 20.0,
        60.1, and 180.4 K). 0.45mm continuum
contours are  $\pm$0.01, 0.03, 0.09, 0.27, and
        0.81 Jy/beam. 
              }
         \label{}
   \end{figure*}

\section{Observations}

We present ALMA observations of the Red Rectangle, using receiver
bands 7, to map the \doce\ and \trece\ \jtd\ transitions, and 9, to map
the CO \jsc\ transition\footnote{ALMA is a partnership of ESO
  (representing its member states), NSF (USA) and NINS (Japan),
  together with NRC (Canada) and NSC and ASIAA (Taiwan), in cooperation
  with the Republic of Chile. The Joint ALMA Observatory is operated by
  ESO, AUI/NRAO and NAOJ. We made use of the ALMA
  dataset  
ADS/JAO.ALMA\#2011.0.00223.S. }.
The backends were set to achieve a spectral resolution of 0.21 \kms.
Band 7 observations were performed in October 2012 and those of band 9
in November; the data were delivered in December 2012. During the
observations, the array consisted of 21 to 24 antennas distributed in a
configuration that extended up to 380 m. RF and flux calibrators were
observed in each run, and two phase calibrators were observed alternatively
with the source.  The delivered calibrated data were significantly
improved with a better-adapted calibration strategy. See more details
in Appendix A.

The image deconvolution was made by using robust weighting (with a
threshold of 1) for band 7 and natural weighting for band 9.
The resulting brightness {\em rms} is 5 mJy/beam in the CO
\jtd\ channel maps, for a spectral resolution of 0.42 \kms\ and a
synthetic beam with a HPBW of 0\farcss 48$\times$0\farcss 44. The {\it
  rms} becomes $\sim$ 7 mJy/beam in the channel maps with the brightest
emission.
For \doce\ \jsc\ maps the {\em rms} is $\sim$ 30 mJy/beam with a
synthetic beam of 0\farcss 31$\times$0\farcss 25 and 0.42
\kms\ resolution.
For the continuum maps, a brightness {\em rms} of 3 mJy/beam is
obtained in band 7 and of 8 mJy/beam in band 9.  The continuum emission
extends in total $\sim$ 3$''$ in the direction of the equatorial disk.
The integrated continuum flux varies, because of the source spectral
index, in the ranges 0.59-0.66 and 3.2-3.6 Jy, respectively for bands 7
and 9.  Calibration was made with the CASA software package, while for
the imaging and data analysis we used GILDAS.  An optical image was
obtained from the HST archive and used for comparison with the CO data,
see details in App.\,A.

\section{Results, simple modeling of our ALMA maps}

Our maps (Figs.\ 1, 2, and 3) show the complex structure of the Red
Rectangle. 
The already known rotating disk in the nebula equator (previously
mapped in \doce\ \juc\ and \jdu\ with lower quality by Bujarrabal et
al.\ 2005) is accurately described in our maps. Together with this
disk, an X-shaped structure delineating the lobes of the optical nebula
(mostly its inner part) are mapped in molecular emission for the first
time. This component must be in expansion, because their emission is
detected at too high velocities and large distances to the center. (The
gravitational attraction is given by the rotation in the disk, which
shows high velocities only very close to the center, as expected for a
Keplerian or Keplerian-like law; see Bujarrabal et al.\ 2005.)

The presence of CO-emitting outflows in the Red Rectangle was suggested
in Paper I, from careful modeling of mm-wave and FIR CO lines; but only
indirect indications of the outflow were provided and qualitative ideas
on its spatial distribution and physical conditions could be deduced.
Our results clearly confirm that prediction. The emission from the
X-shaped component is indeed 
dominant in the \doce\ \jsc\ line, but is rather weak in the
\trece\ \jtd\ one.

\doce\ and \trece\ \jtd\ show a similar peak brightness in the disk,
with a remarkably high value not much lower than the expected kinetic
temperatures ($\sim$ 100 K, Paper I), showing that, in the central
regions, these lines are mostly thermalized and optically thick. The
peak brightness of \doce\ \jsc\ is also comparable, but its intensity
significantly decreases toward the end of the disk, indicating that its
excitation is relatively low in the outer disk. The outflowing
component is very bright in \doce\ \jsc, notably at distances smaller
than $\sim$2$''$ from the equator. \doce\ \jtd\ is about a factor two
weaker in these regions, and \trece\ \jtd\ is much weaker, almost
undetected. This shows that the observed lines are not opaque in the
outflow and that its temperature is relatively high (\gsim\ 100 K).

   \begin{figure*}
   \centering
   \includegraphics[width=17.3cm]{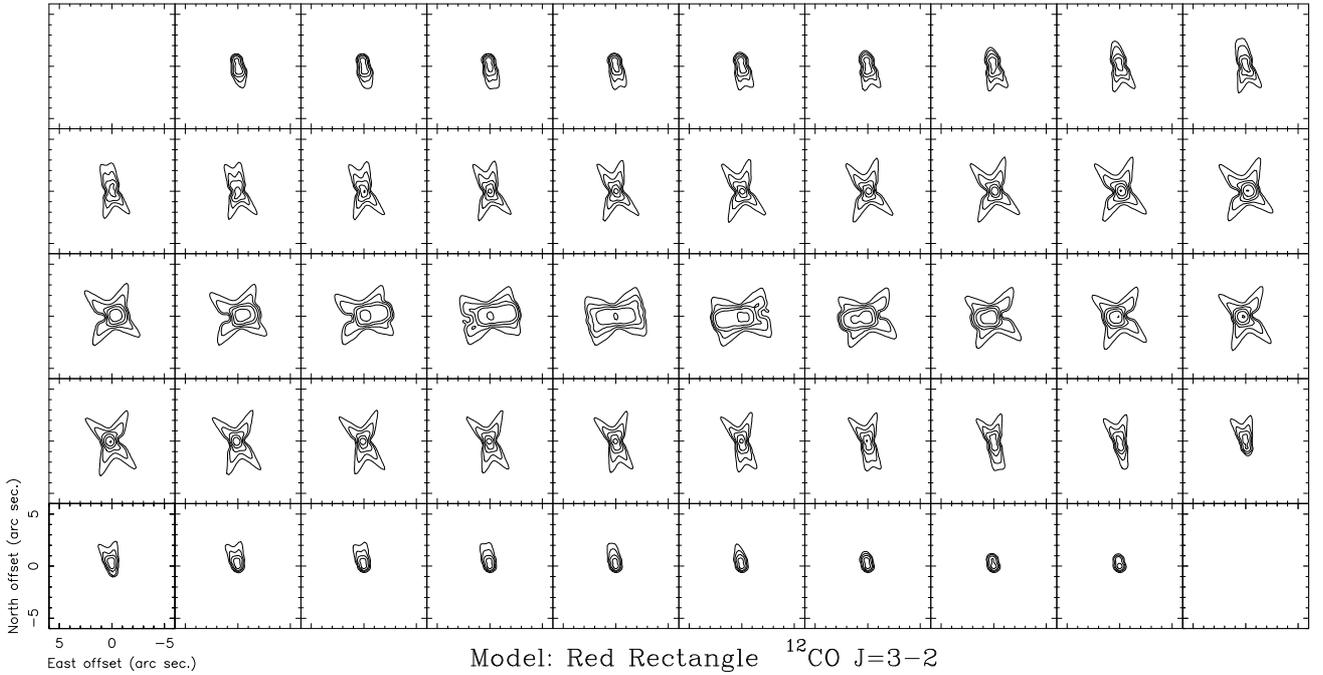}
      \caption{Calculations from our best-fit simple model for the
        \doce\ \jtd\ line emission. The contours and angular and
        velocity units are the same as for 
        Fig.\ 1, which shows the observational data.  }
         \label{}
   \end{figure*}

We have performed a simple model fitting of our \doce\ \jtd\ maps. We
used a code very similar to that described by Bujarrabal et
al.\ (2005). In this code, the excitation is described by an LTE
approximation, which is valid for low-$J$ high-opacity transitions,
that are easily thermalized, but certainly not in general (Paper
I). For that reason we applied this treatment only to \doce\ \jtd\ and
here we just give a brief discussion on the modeling of our ALMA
results. A more detailed treatment of the line excitation, including
model fitting for all lines, requires complex calculations using an
'exact' non-local treatment in two dimensions (Paper I) and a very
careful and detailed discussion of the results, which is clearly out of
the scope of this {\em letter} and is delayed to a forthcoming paper.

We adopted the same disk model as that deduced in Paper I (where all
parameters are discussed in detail). We only decreased the disk density
by factor 2, because the predicted line emission from this component
was stronger than observed. We introduced a new component in expansion,
which is kept as simple as possible and symmetric with respect to
the polar axis and equatorial plane. The results of our modeling, in
the form of synthetic maps of \doce\ \jtd\ in the same units as the
observational data, are shown in Fig.\ 4. The shape, dynamics, and
density in the best-fit model nebula are depicted in Fig.\ 5; the
temperature of the outflow is assumed to be constant and equal to 200
K. In Fig.\ 5, we represent parameters in a plane perpendicular to the
equator, therefore the disk is seen edge on and the rotational
velocity, which is the same as in Paper I, cannot be displayed.

As we see, the model prediction is satisfactory.
The emission of the disk is well reproduced assuming practically the
same structure and conditions as those derived in Bujarrabal et
al.\ (2005) and Paper I. The emission from the expanding gas (at
relatively high velocity shifts) is also well explained. It is
remarkable that the properties of the
outflow are roughly compatible with those tentatively deduced in Paper
I to explain the profiles of high-$J$ CO lines.  However, several
caveats must be kept in mind. First of all, the nebula is not exactly
symmetric with respect to the equator and the axis. This cannot be
reproduced by our model, which tends to give a
kind of average. Second, we assumed a disk density 
somewhat lower than previously thought; this may result from the
contribution of the outflow to the total emission and our LTE
assumption, which would not account for some underexcitation in this
line (since the density is not very high). We recall that a more
detailed modeling must include the fitting of our \trece\ \jtd\ and
\doce\ \jsc\ maps and much more complex and difficult calculations,
and is deferred to a forthcoming work. Our modeling is preliminary,
but we think that it yields the main results that can be extracted from
our data and a reasonable description of the nebula.

\section{Conclusions} 

We present high-quality ALMA observations of \doce\ and
\trece\ \jtd\ and \doce\ \jsc\ line emission in the Red Rectangle, see
Figs.\ 1, 2, and, 3. In these figures we also show the maps of the
submm-wave continuum, as well as an HST image of the nebula for
comparison (Sect.\ 2).  We performed a preliminary modeling of the
\doce\ \jtd\ line (which is the easiest to model, Sect.\ 3), which
yielded satisfactory predictions; our best-fit synthetic maps and
deduced nebula properties are shown in Figs.\ 4 and 5.
Our main results can be summarized as follows:

   \begin{list}{}{\topsep 0mm \itemsep 0mm \parskip 0mm \leftmargin 4mm
    \partopsep -0mm \parsep 0mm \listparindent 0mm \itemindent -4mm}

\item{\bf 1:} The Red Rectangle is a complex nebula, showing both a
  rotating disk and a component in expansion. The coexistence of
  extended rotating disks
  and outflows is probably common in post-AGB objects, at least in
  low-mass nebulae surrounding binary stars (Sect.\ 1, Paper II).

      \item{\bf 2:} The high brightness measured in the central panels
        suggests opaque lines in the disk. \doce\ \jsc\ emission is
        particularly intense in the outflow, indicating relatively high
        excitation and low opacities in this component.

      \item{\bf 3:} Our interpretation of the data is compatible with
        the properties of the disk deduced in previous works.

      \item{\bf 4:} The gas in expansion basically occupies the region
        placed within the disk and the double cone delineated by the
        X-shaped optical images. The optical nebula must then
        correspond to the inner illuminated region of the expanding
        lobes, the density within the double cone being much lower. The
        velocity of the outflow is moderate, $\sim$ 10 \kms, with a
        trend to decrease in regions close to the equator. The density
        of this expanding component increases toward the center, with
        values between 10$^3$ and 10$^5$ cm$^{-3}$; the temperature of
        the gas is kept constant at 200 K.

   \end{list}

\begin{acknowledgements}
This work has been supported by the Spanish MICINN, program CONSOLIDER
INGENIO 2010, grant ``ASTROMOL" (CSD2009-00038). CSC is partially
supported by the Spanish MINECO, grants AYA2009-07304 and
AYA2012-32032. We used the SIMBAD and MAST databases.
\end{acknowledgements}

   \begin{figure}
   \centering
   \includegraphics[width=6.5cm,angle=0]{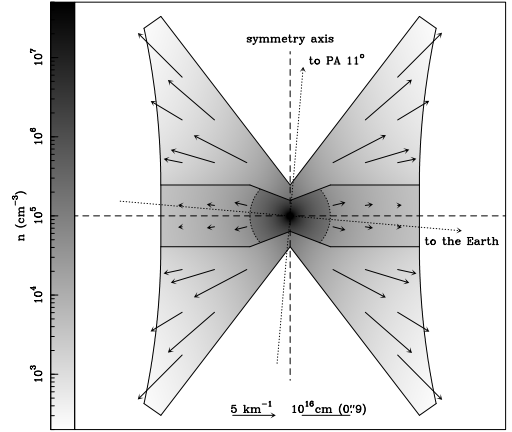}
      \caption{Structure, velocity, and density distribution in our
        best-fit model disk and outflow. We represent parameters for a
        plane perpendicular to the equator, i.e.\ the disk is seen
        edge-on and only expansion is shown.
}
         \label{}
   \end{figure}


\newpage
\appendix

\section{Further details on observations and data reduction}

Two observing runs were carried out in band 7 in October 21 and 22,
2012, of 80 and 100 min respectively, with which we obtained a total of
71 min of correlations on source.  The band 9 observations were
performed in November 3 and 4, 2012, in four runs with durations of 133,
127, 64, and 111 min, which allowed accumulating a total of 135 min of
correlations on source.  For band 7, J0522-364 was observed for the RF
calibration.  For band 9, J0538-440 or 3\,C84 were observed instead.
Callisto was always observed for absolute flux calibration, but for one
run in band 9 in which data in Ceres was obtained.  J0607-085 and
J0609-157 were observed every 10 to 20 min for gain calibration.

ALMA staff provided first calibrated data in December 2012, which were
considerably improved by us. In particular, we made sure that the
absorption CO lines detected at the position of the RF calibrator did
not get into source data. The bandpass was calibrated by fitting cubic
splines.  Also, having identified different RF solutions for the
different observed runs, an independent calibration was performed per
track. For gain calibration, data from the two observed calibrators
were considered.  Later, from these first calibrated data,
selfcalibration was carried out by using the bright and moderately
extended continuum emission of the Red Rectangle.  All those processes
resulted in a very significant improvement in the quality of final
images.

The flux calibration was based on the Butler-JPL-Horizons 2010 model
for solar system bodies. A recent upgrade in that model (including
e.g.\ Herschel data) suggests that the intensity here presented could
be slightly overestimated, by 10\% in band 7, and by 5-10\% in band 9.

In band 7, the total observed frequency ranges were 330.354-330.822,
333.152-333.620, 345.106-345.574, and 345.562-346.030 GHz, aiming to
map mainly the emission of the \doce\ and
\trece\ \jtd\ transitions.  In band 9, we observed the intervals
673.540-674.477, 677.040-677.977, 687.805-688.742 and 691.004-691.941
GHz to image the \doce\ \jsc\ line.  

A WFPC2/f622w optical image was obtained from the HST archive. The
optical and ALMA images were aligned adopting the astrometry provided
by the Hubble Legacy Archive, which we have checked is reliable using
the 2MASS coordinates of the 11 field stars detected within the WFPC2
field, and correcting for the Red Rectangle's proper motion (as
measured by Hipparcos and listed in the SIMBAD database, $\delta_{\rm
  RA}$=$-$6.46$\pm$2.21 and $\delta_{\rm Dec}$=$-$22.74$\pm$2.03
mas\,yr$^{-1}$, van Leeuwen 2007) for a time difference of 13.3\,yr
between the HST and ALMA observations.  The resulting alignment of the
centroids, within \lsim\ 0\farcss 1, is found to be satisfactory, in
view of the uncertainty in the stellar positions and proper movements.

\end{document}